\documentclass[12pt,a4]{article}

\usepackage{amsthm,amsmath,latexsym,amssymb,amsfonts}
\usepackage{graphicx,lscape,fancyhdr,array,stmaryrd}

\newcommand{\beq}{\begin{equation}}
\newcommand{\eeq}{\end{equation}}
\newcommand{\bee}{\begin{eqnarray}}
\newcommand{\eee}{\end{eqnarray}}
\newcommand{\bml}{\begin{multline}}
\newcommand{\eml}{\end{multline}}
\newcommand{\bsplit}{\begin{split}}
\newcommand{\esplit}{\end{split}}

\newcommand{\pl}{\partial}

\newcommand{\mcc}{\ensuremath{E}}

\newcommand{\iso}{{\ensuremath{iso(d-1,1)}}}
\newcommand{\lorentz}{{\ensuremath{o(d-1,1)}}}
\newcommand{\msv}{{\ensuremath{o(d-1)}}}
\newcommand{\mls}{{\ensuremath{o(d-2)}}}

\newcommand{\field}[4]{{#1}_{{#2}_{#3}...{#2}_{#4}}}

\newcommand{\phif}{\field{\phi}{\mu}{1}{s}}

\newcommand{\smf}{{\textstyle\frac{s(s-1)}2}}

\newcommand{\phitr}[2]{{{\phi'}_{#1}^{#2}}}

\newcommand{\tr}[2]{{#1}^{#2}_{\phantom{#2}#2}}
\newcommand{\crd}{{Q}}
\newcommand{\mtm}{{\Pi}}
\newcommand{\Ham}{{H}}
\newcommand{\NN}{{d-2}}

\begin{document}

\begin{flushright}
\vspace{1mm}
FIAN/TD/14-06\\
\end{flushright}

\vspace{1cm}

\begin{center}
{\bf \Large  Transverse Invariant Higher Spin Fields} \vspace{1cm}

\textsc{E.D. Skvortsov\footnote{skvortsov@lpi.ru} and M.A.
Vasiliev\footnote{vasiliev@lpi.ru}}

\vspace{.7cm}

{\em I.E.Tamm Department of Theoretical Physics, P.N.Lebedev
Physical
Institute,\\
Leninsky prospect 53, 119991, Moscow, Russia}

\vspace{3mm}

\end{center}

\vspace{1cm}
\begin{abstract}
It is shown that a symmetric massless bosonic higher-spin field
can be described by a traceless tensor field with reduced
(transverse) gauge invariance. The Hamiltonian analysis of the
transverse gauge invariant higher-spin models is used to control a
number of degrees of freedom.
\end{abstract}

\section{Introduction}
\setcounter{equation}{0}

It was shown recently in \cite{Alvarez:2006uu}, \cite{Blas:2007pp}
that a free massless spin two field (i.e. linearized gravity) can
be consistently described by a traceless rank-$2$ tensor field
with transverse gauge symmetry that corresponds to linearized
volume-preserving diffeomorphisms. We extend this result to
massless fields of arbitrary spin by showing that a spin-$s$
symmetric massless field can be described by a rank-$s$ traceless
symmetric tensor. This formulation is in some sense opposite to
the approach developed in
\cite{Francia:2007qt,Bekaert:2003az,deMedeiros:2003dc} where a
massless field is described by a traceful tensor. Recall that the
standard Fronsdal's formulation of a spin-$s$ massless field
operates with a rank-$s$ double traceless tensor
\cite{Fronsdal:1978rb}. For recent reviews on higher-spin (HS)
gauge theories see \cite{Solvay}.

Although, like in the case of gravity, the obtained model is a
gauge fixed version of the original Fronsdal model
\cite{Fronsdal:1978rb} the equivalence is not completely trivial.
Actually, the standard counting of degrees of freedom is that each
gauge parameter in the gauge transformations with first order
derivatives kills two degrees of freedom \cite{Gitman:1990qh}.
Therefore one can expect that the invariance under reduced gauge
symmetry may be not sufficient to compensate all extra degrees of
freedom. As we show this is not the case. The reason is that the
remaining gauge symmetry parameters satisfy the differential
transversality conditions $\pl^{\nu}
\xi_{\nu\mu_2...\mu_{s-1}}=0$.

Generally, as explained in this paper, a partial gauge fixing at
the Lagrangian level can give rise to a model which, if treated
independently of the original gauge model, may differ from the
latter. In particular, the Hamiltonian interpretation of the gauge
fixed Lagrangian model may differ from that of the original model.
This can happen in the case where the gauges and constraints on
gauge parameters are differential. For example, as shown in
Section 5, this does happen in electrodynamics in the temporary
gauge. Since the transversality condition on the gauge parameter
is also of this type, a more careful analysis of the counting of
the number of degrees of freedom in the model under consideration
is needed. The Hamiltonian analysis of Section
\ref{HamiltionianAnalysis} shows that the transverse gauge
invariant
 HS model has as
many degrees of freedom as the original Fronsdal model in the
topologically trivial situation.

Note, that the original Lagrangian and field/gauge
transformations content for a massless  field of an
arbitrary spin were derived by Fronsdal in \cite{Fronsdal:1978rb}
by taking the zero rest mass limit $m^2\rightarrow0$
in the Lagrangian of Singh and Hagen for a massive HS
field of \cite{Singh:1974qz}. To the best of our knowledge,
it has not been analyzed in the literature what is a minimal
field content appropriate for the description a massless HS field.
 The proposed formulation operates in
terms of an irreducible Lorentz tensor field, thus being minimal.
It is equivalent to the Fronsdal's one in the topologically
trivial situation but may differ otherwise. Also let us note that
since it has a relaxed gauge symmetry compared to that of the
Fronsdal formulation, it may in principle have more freedom at the
interaction level.

The layout of the rest of the paper is as follows. In Section
\ref{FreeMasslessHSFields} we recall the standard description of
massive and massless fields of arbitrary spin. In Section
\ref{TransverseWeylHSFields}, transverse and Weyl invariant
Lagrangian is constructed and a generating action is given. The
equivalence of transverse and Weyl invariant Lagrangian to the
Fronsdal's Lagrangian is checked in Section
\ref{SpectrumUnitarity}. Hamiltonian analysis and examples are
given in Section \ref{HamiltionianAnalysis}.
\section{Free Massless Higher-Spin Fields}\label{FreeMasslessHSFields}
\setcounter{equation}{0} A spin-$s$  bosonic totally symmetric
massive field in Minkowski space  can be described on shell
\cite{Dirac:1936} by a totally symmetric tensor field
$\varphi_{\mu_1...\mu_s}$\footnote{Greek indices
$\mu$,$\nu$,$\lambda$,$\rho=0,...,d-1$ are vector indices of
$d$-dimensional Lorentz algebra \lorentz.
$\pl_\mu\equiv\frac{\pl}{\pl x^\mu}$, $\square\equiv \pl^\nu
\pl_\nu$ and indices are raised and lowered by mostly minus
invariant tensor $\eta_{\mu\nu}$ of \lorentz. A group of indices
to be symmetrized is denoted by placing them in brackets or,
shortly, by the same letter. For example,
$\pl_\mu\phi_\mu\equiv\pl_{(\mu_1}\phi_{\mu_2)}
\equiv\frac12\left(\pl_{\mu_1}\phi_{\mu_2}+\pl_{\mu_2}\phi_{\mu_1}\right)$.}
that satisfies the conditions
\beq\label{MassiveEqs}\begin{split}
&(\square+m^2)\varphi_{\mu_1...\mu_s}=0, \\
&\pl^\nu\varphi_{\nu\mu_2...\mu_s}=0, \\
&\varphi^{\nu}_{\phantom{\nu}\nu\mu_3...\mu_s}=0.
\end{split}\eeq
These form the complete set of local Poincare-invariant conditions
on $\varphi_{\mu_1...\mu_s}$.  In the massless case  $m^2=0$
a gauge invariance with an on-shell traceless rank-$(s-1)$ tensor gauge
parameter reduces further the number of physical degrees of freedom.

As pointed out by Fierz and Pauli in \cite{Fierz:1939ix}, for
(\ref{MassiveEqs}) to be derivable from a Lagrangian a set of
auxiliary fields has to be added for $s>1$ (in the case of spin
two considered by Fierz and Pauli this is a scalar auxiliary field
$\varphi$, which together with a traceless $\varphi_{\mu_1\mu_2}$
forms a traceful field
$\phi_{\mu_1\mu_2}=\varphi_{\mu_1\mu_2}+\eta_{\mu_1\mu_2}\varphi$).
Auxiliary fields are zero on shell, thus carrying no physical
degrees of freedom. For totally symmetric massive fields of
integer spins, the Lagrangian formulation with a minimal set of
auxiliary fields was worked out by Singh and Hagen in
\cite{Singh:1974qz}. For a spin-$s$ field they introduced a set of
auxiliary fields, which consists of symmetric traceless tensors of
ranks $s-2, s-3,\ldots 0$. An elegant gauge invariant
(Stueckelberg) formulation was proposed by Zinoviev in
\cite{Zinoviev}. (For alternative approaches to massive fields see
also \cite{Bekaert:2003uc,Metsaev:2005ar,Buchbinder:2006ge} and
references therein.)

The Lagrangian of a spin-$s$ massless field can be obtained
\cite{Fronsdal:1978rb} in the limit $m^2\rightarrow0$. The
auxiliary fields of ranks from $0$ to $(s-3)$ decouple while the
residual rank-$(s-2)$ traceless auxiliary field
$\varphi_{\mu_1...\mu_{s-2}}$ and the physical rank-$s$ traceless
field $\varphi_{\mu_1...\mu_s}$ form the symmetric field
$\phi_{\mu_1...\mu_s}=\varphi_{\mu_1...\mu_s}+\eta_{(\mu_1\mu_2}
\varphi_{\mu_3...\mu_{s-2})}$ that satisfies the double
tracelessness condition \beq\label{Doubletracelessness}
\eta^{\mu_1\mu_2}\eta^{\mu_3\mu_4}\phi_{\mu_1...\mu_s}=0, \eeq
which makes sense for $s\geq4$. The resulting Lagrangian possesses
gauge invariance with a traceless rank-$(s-1)$ gauge parameter
$\xi_{\mu_1...\mu_{s-1}}$, \beq\label{MasslessGauge}\delta
\phif=s\pl_{(\mu_1}\xi_{\mu_2...\mu_s)},\qquad
\xi^{\nu}_{\phantom{\nu}\nu\mu_3...\mu_{s-1}}=0. \eeq In the spin
two case of  linearized gravity, the gauge law
(\ref{MasslessGauge}) corresponds to linearized diffeomorphisms.

Let us write down a most general bilinear action and Lagrangian
(modulo total derivatives) of a double traceless field with at
most two derivatives as \beq\label{Lagrangian}
\mathcal{L}=(-)^s\sum_{\alpha=a,b,c,f,g}\mathcal{L}_{\alpha},\qquad\qquad
S=\int d^dx \mathcal{L},\eeq where \beq\begin{split}
&\mathcal{L}_a=\frac{a}2\pl_\nu\phi_{\mu_1...\mu_s}\pl^\nu\phi^{\mu_1...\mu_s},\\
&\mathcal{L}_b=-\frac{bs(s-1)}4\pl_\nu\phi^{\rho}_{\phantom{\rho}\rho\mu_3...\mu_s}\pl^\nu\phi_{\lambda}^{\phantom{\lambda}\lambda\mu_3...\mu_s},\\
&\mathcal{L}_c=-\frac{cs}2\pl^\nu\phi_{\nu\mu_2...\mu_s}\pl_\rho\phi^{\rho\mu_2...\mu_s},\\
&\mathcal{L}_f=\frac{fs(s-1)}2\pl_\nu\phi^{\rho}_{\phantom{\rho}\rho\mu_3...\mu_s}\pl_\lambda\phi^{\lambda\nu\mu_3...\mu_s},\\
&\mathcal{L}_g=-\frac{g
s(s-1)(s-2)}8\pl^\nu\phi^{\rho}_{\phantom{\rho}\rho\nu\mu_4...\mu_s}\pl_\lambda\phi_{\sigma}^{\phantom{\sigma}\sigma\lambda\mu_4...\mu_s}
\end{split}\eeq with arbitrary coefficients $a,$ $b,$ $c,$ $f,$ $g$.
For $\mathcal{L}$ to describe a spin-$s$ field, the coefficient $a$ has to be
nonzero (so, we set $a=1$).

The variation of (\ref{Lagrangian})
is
\beq
\delta\mathcal{L}=\left(G_{\mu_1...\mu_s}-\frac{s(s-1)}{2(\Upsilon-2)}\eta_{(\mu_1\mu_2}G^\rho_{\phantom{\rho}\rho\mu_3...\mu_s)}\right)\delta\phi^{\mu_1...\mu_s},
\eeq where $\Upsilon=d+2s-4$ and \beq\begin{split}
& G_{\mu_1...\mu_s} = \square \phi_{\mu(s)}-b
\smf\eta_{\mu\mu}\square\phi^{\lambda}_{\phantom{\lambda}\lambda\mu(s-2)}-cs\pl_\mu
\pl^\nu\phi_{\nu\mu(s-1)}+\\+&f\smf\left(\eta_{\mu\mu}\pl^\nu\pl^\lambda\phi_{\nu\lambda\mu(s-2)}+
\pl_\mu\pl_\mu\phi^{\lambda}_{\phantom{\lambda}\lambda\mu(s-2)}\right)-
g{\textstyle\frac{s(s-1)(s-2)}4} \eta_{\mu\mu}\pl_\mu
\pl^\nu\phi^{\lambda}_{\phantom{\lambda}\lambda\nu\mu(s-3)}\,.
\end{split}\eeq
The requirement that the action is invariant under
(\ref{MasslessGauge}) fixes the coefficients $a=b=c=f=g$
\cite{Curtright:1979uz}.

\section{Transverse and Weyl Invariant Massless Higher-Spin
Fields}\setcounter{equation}{0} \label{TransverseWeylHSFields}

Let us consider a weaker condition on the action imposed by the
reduced gauge symmetry (\ref{MasslessGauge}) with the transverse
gauge parameter $\xi_{\mu_1...\mu_{s-1}}$

\beq\label{TDiffGauge}
\delta
\phif=s\pl_{(\mu_1}\xi_{\mu_2...\mu_s)},\qquad \pl^{\nu}
\xi_{\nu\mu_2...\mu_{s-1}}=0,\qquad
\xi^{\nu}_{\phantom{\nu}\nu\mu_3...\mu_{s-1}}=0.\eeq

The invariance of action (\ref{Lagrangian}) under
(\ref{TDiffGauge}) fixes only the ratio $a/c=1$ while the rest of
the coefficients remains free. This ambiguity can be used to look
for another symmetry to kill extra degrees of freedom. Taking into
account the double tracelessness condition
(\ref{Doubletracelessness}), a use of rank-$(s-2)$ symmetric
traceless gauge parameter $\zeta_{\mu_1...\mu_{s-2}}$ is a natural
option

\beq \label{WeylGauge}\delta
\phif=\smf\eta_{(\mu_1\mu_2}\zeta_{\mu_3...\mu_s)},\qquad
\zeta^{\nu}_{\phantom{\nu}\nu\mu_3...\mu_{s-2}}=0.\eeq

The requirement for (\ref{Lagrangian}) to be invariant under the
additional (Weyl) symmetry (\ref{WeylGauge}) fixes the rest of the
coefficients \beq b=\frac{\Upsilon+2}{\Upsilon^2},\qquad
 f=\frac{2}{\Upsilon},\qquad
g=\frac{-2(\Upsilon-4)}{\Upsilon^2}.
\eeq

Note that, not too surprisingly, the resulting Lagrangian
(\ref{Lagrangian}) can be obtained from the Fronsdal's Lagrangian
(i.e. that with $a=b=c=f=g=1$) via the substitution \beq
\tilde{\phi}_{\mu_1...\mu_s}=\phif-\frac1{\Upsilon}\smf\eta_{(\mu_1\mu_2}\phi^{\nu}_{\phantom{\nu}\nu\mu_3...\mu_s)},\qquad\tilde{\phi}^{\nu}_{\phantom{\nu}\nu\mu_3...\mu_s}=0.
\eeq

There is a generating action $S^{gen}$ that gives rise both to the
Fronsdal and to the Weyl invariant actions in particular gauges.
$S^{gen}$ results from the Fronsdal action by introducing  a
traceless Stueckelberg field $\chi_{\mu_1...\mu_{s-2}}$ of
rank-$(s-2)$ via the substitution \beq
\phi_{\mu_1...\mu_s}\longrightarrow\phi_{\mu_1...\mu_s}+\smf\eta_{(\mu_1\mu_2}\chi_{\mu_3...\mu_s)},
\eeq where $\phi_{\mu_1...\mu_s}$ is the double traceless field. The
gauge transformations are \beq \delta
\phif=s\pl_{(\mu_1}\xi_{\mu_2...\mu_s)}+\smf\eta_{(\mu_1\mu_2}\varepsilon_{\mu_3...\mu_s)},\qquad
\delta
\chi_{\mu_1...\mu_{s-2}}=-\varepsilon_{\mu_1...\mu_{s-2}}\eeq with
$\varepsilon_{\mu_1...\mu_{s-2}}$ being a traceless rank-$(s-2)$
gauge parameter. Fixing $\chi_{\mu_1...\mu_{s-2}}$ to zero by the
gauge parameter $\varepsilon_{\mu_1...\mu_{s-2}}$, we  obtain the
spin-$s$ Fronsdal's Lagrangian. Alternatively, we can gauge fix
the trace of $\phif$ to zero by the same Stueckelberg parameter
$\varepsilon_{\mu_1...\mu_{s-2}}$. The leftover symmetry is with
\beq
\varepsilon_{\mu_1...\mu_{s-2}}=
\frac{\Upsilon}2\pl^\nu\xi_{\nu\mu_1...\mu_{s-2}}.
\eeq
Then, gauge fixing the field $\chi_{\mu_1...\mu_{s-2}}$ to zero
gives the Lagrangian (\ref{TrAction}) and constraint
\beq\label{divLessCovConstr} \pl^{\nu} \xi_{\nu\mu_2...\mu_{s-1}}=0.\eeq
Thus $S^{gen}$ reduces to the
transversely invariant action (\ref{TrAction}) and Fronsdal action
in particular gauges. Note that a generating action of this type
naturally appears in the BRST analysis as discussed by Pashnev and
Tsulaia in \cite{Pashnev:1998ti}.

Now we are in a position to check whether this theory is unitary and
describes the correct number of physical degrees of freedom of a
spin-$s$ massless representation of \iso, thus being equivalent to
the conventional spin-$s$ Fronsdal massless theory.

\section{Spectrum}\label{SpectrumUnitarity}
\setcounter{equation}{0} Having fixed pure algebraic gauge
symmetry with parameter $\zeta_{\mu_1...\mu_{s-2}}$ to eliminate
the trace of  $\phi_{\mu_1...\mu_s}$ one gets the Lagrangian
\beq\label{TrAction}
\mathcal{L}=(-)^s\left(\frac{1}2\pl_\nu\phi_{\mu_1...\mu_s}\pl^\nu\phi^{\mu_1...\mu_s}-
\frac{s}2\pl^\nu\phi_{\nu\mu_2...\mu_s}\pl_\rho\phi^{\rho\mu_2...\mu_s}\right)
\eeq
with, respectively, the equations of motion, gauge transformation law and constraints
\beq\label{TDiffEqs}\begin{split}
&\square\phi_{\mu_1...\mu_s}-s\pl_{(\mu_1}\pl^\nu\phi_{\nu\mu_2...\mu_s)}+\frac{s(s-1)}{\Upsilon}
\eta_{(\mu_1\mu_2}\pl^{\nu}\pl^{\rho}\phi_{\nu\rho\mu_3...\mu_s)}=0,
\\ &\delta \phif=s\pl_{(\mu_1}\xi_{\mu_2...\mu_s)},\\ & \pl^{\nu}
\xi_{\nu\mu_2...\mu_{s-1}}=0,\qquad
\xi^{\nu}_{\phantom{\nu}\nu\mu_3...\mu_{s-1}}=0,\qquad
\phi^{\nu}_{\phantom{\nu}\nu\mu_3...\mu_s}=0.\end{split}\eeq

To analyze the physical meaning of these equations and gauge transformations it is convenient to use the standard
momentum frame \beq p_\mu=(\mcc/\sqrt{2},0,...,0,\mcc/\sqrt{2}),\qquad p^\nu p_\nu=0
\eeq
and light-cone
coordinates\footnote{Lower case Latin indices $i,j,...$ are vector indices of \mls. The corresponding invariant
metric is $\delta_{ij}=\mbox{diag}(\overbrace{+...+}^{d-2})$.} \beq x^{\pm}=(x^0\pm x^d)/\sqrt{2},\quad x^i -
\mbox{unchanged}, \eeq in which the metric $\eta_{\mu\nu}$ has the form  \beq \eta_{+-}=\eta_{-+}=1,\quad
\eta_{ij}=-\delta_{ij}\,. \eeq

We use the following notation
for components of $\phif$ \beq
\phi_{+(k),-(m),i(s-k-m)}\equiv\phi_{\underbrace{\scriptstyle
+...+}_k \underbrace{\scriptstyle -...-}_m i_1 ... i_{s-k-m}}.\eeq
$\phitr{+(k),-(m),i(s-k-m-2)}{}$ denotes the trace
$\phi_{+(k),-(m),i(s-k-m-2)jl}\delta^{jl}$ of \mls\ indices.

The system (\ref{TDiffEqs}) reduces to
\beq\label{TDiffEqsLightCone}\begin{split}
&k(1-{\textstyle\frac{2m}{\Upsilon}})\phi_{+(k-1),
-(m+1),i(s-k-m)}+{\textstyle\frac{(s-k-m)(s-k-m-1)}{\Upsilon}}\delta_{ii}\phi_{+(k),
-(m+2),i(s-k-m-2)}=0,
\\
&\delta\phi_{+(k), -(m),i(s-k-m) }=k\xi_{+(k-1), -(m),i(s-k-m)}, \\
& \xi_{+(k), -(m>0),i(s-k-m)}=0,\qquad 2\xi_{+(k+1),
-(m+1),i(s-k-m-2)}=\xi_{+(k), -(m),i(s-k-m-2)}',\\& 2\phi_{+(k+1),
-(m+1),i(s-k-m-2)}=\phi_{+(k), -(m),i(s-k-m-2)}'.\end{split}\eeq
The first equation of (\ref{TDiffEqsLightCone}) implies that
$\phi_{+(k-1), -(m+1),i(s-k-m)}$ is a pure trace for
$m=0...(s-1)$, $k=1...(s-m)$. As a result the on shell non-zero
components are \mls\
traceless components
$\phi_{+(k),-(0),i(s-k)}$, $k=0...s$. However, those with
$k=1...s$ are pure gauge. Thus, only the traceless component of
$\phi_{+(0),-(0),i(s)}\equiv\phi_{i(s)}$ is physical, describing a
spin-$s$ symmetric representation of the massless little group
\mls. Unitarity of the theory  follows from the equivalence
of the transverse-invariant and Fronsdal's Lagrangians in the
sector of physical degrees of freedom.

A  less trivial question not answered by this analysis is whether
the leftover gauge symmetries in a partially gauge fixed model
remain gauge symmetries of the latter model treated independently
(say, if the original model was not known). Complete answer to
this question is provided by the Hamiltonian analysis. To
illustrate what could happen let us start with the spin one
example.

\section{Hamiltonian analysis}\setcounter{equation}{0}\label{HamiltionianAnalysis}

\subsection{Example of spin one in the temporary gauge}\label{Spin One Example}
An instructive example is provided by Maxwell electrodynamics formulated in terms of a gauge potential $A_\mu$
\beq\label{MaxwellLagrangian}\begin{split} &\mathcal{L}=\frac12\left(\pl^\mu A_\mu\pl^\nu A_\nu - \pl_\mu
A_\nu\pl^\mu A^\nu\right),\\ & \square A_\mu-\pl_\mu \pl^\nu A_\nu=0, \\ &\delta A_\mu=\pl_\mu \xi(x).
\end{split}\eeq

Imposing the temporary gauge $A_0=0$ at the Lagrangian level we obtain the gauge fixed Lagrangian\footnote{Capital
Latin indices $I,J,K,...$  are vector indices of \msv, e.g. $\mu$=(0,I). Dot denotes the time derivative, i.e.
$\dot{\phi}\equiv\pl_0\phi$ } \beq \label{L0} \mathcal{L}=\frac12\left(\dot{A}_K\dot{A}^K+\pl^I A_I\pl^J A_J-\pl_I
A_J\pl^I A^J\,\right).
\eeq
Expressing all velocities via momenta we arrive at the unconstrained dynamics with the
Hamiltonian \beq \label{H0} \Ham=\frac12\left(\mtm_K\mtm^K+\pl_I A_J\pl^I A^J-\pl^I A_I\pl^J A_J\right)\,. \eeq

Clearly, the gauge fixed Lagrangian (\ref{L0}) describes $d-1$
degrees of freedom ($2(d-1)$ in the phase space). This is to be compared
with the $d-2$ degrees of freedom (2$(d-2)$ in the phase space) of the original
model. This mismatch has the following origin. One additional
phase space degree of freedom comes from the leftover gauge symmetry
parameter that solves
\beq\label{de}
\pl_0 \xi=0\,.
\eeq
Another one is due to the loss of the Gauss law constraint in the theory.

Indeed, in electrodynamics,
the Gauss law $div E = 0$ results from the variation of the action
(\ref{MaxwellLagrangian}) over $A_0$ (for simplicity we set electric current
equal to zero). This equation is lost in the gauge fixed theory (\ref{L0}).
From the gauge invariance a weaker condition follows
\beq
\label{0div}
\pl_0 div E =0.
\eeq
In the dynamical system  (\ref{L0}), the equation (\ref{0div}) is indeed
one of the field equations. But it is not a constraint any more, thus bringing
another phase space degree of freedom into the game.

The naive equivalence argument might be that once some gauge is
reachable by a gauge transformation it can be imposed at the
Lagrangian level because any variation over a gauge fixed variable
can be expressed as a combination of a gauge symmetry variation
and a local variation of the unfixed variables. Generically, this
argument is wrong however because it neglects the issue of
locality. Namely it is not guaranteed that the compensating gauge
symmetry transformation is local in terms of the variation of the
gauge fixed variable because it may require resolution of some
differential equation on the gauge symmetry parameter with respect
to the time variable. In our example, this is manifested by the
condition (\ref{de}). This is why the Gauss law in electrodynamics
is not reproduced in the model (\ref{L0}) treated independently of
the underlying model from which it has been derived.

The conclusion is that, if treated
independently, a gauge fixed model
 (i.e. forgetting the symmetries and field equations of the original model) is
guaranteed to be equivalent to the original one in the gauges that
 impose algebraic (i.e., free of time derivatives) constraints on the
gauge symmetry parameters. This is the case of Stueckelberg
fields and gauge symmetries.

The situation with the spin two field considered in
\cite{Alvarez:2006uu,Blas:2007pp} and with partially fixed HS
gauge fields discussed in this paper is somewhat analogous to the
temporary gauge example discussed in this section because it
involves the differential constraint (\ref{TDiffGauge}) on the
gauge symmetry parameter. It is therefore instructive to reanalyze
the models by the Hamiltonian methods.

\subsection{Spin two}

Let us consider the spin two case in more detail. {(The
Hamiltonian analysis of nonlinear gravity was originally given in
\cite{dirham} (see also the textbook \cite{Gitman:1990qh}))}. The
gauge fixed Lagrangian is \beq
\mathcal{L}=\frac12\left(\pl_\mu\phi_{\nu\lambda}\pl^\mu\phi^{\nu\lambda}-
2\pl^\mu\phi_{\mu\nu}\pl_\lambda\phi^{\lambda\nu}\right),\quad
\tr{\phi}{\nu}=0. \eeq Using notation
$\crd_{IK}\equiv{\phi}_{IK}$, $\crd_I\equiv{\phi}_{0I}$, the
corresponding momenta are
\beq\begin{split}&\mtm^{IJ}=\dot{\crd}^{IJ}- \delta^{IJ}\tr{\dot{\crd}}{K}+2\delta^{IJ}\pl^K \crd_K, \\
&\mtm^I=-2\pl_K {\crd}^{KI}.\end{split}\eeq

As velocities $\dot{\crd}^I$ do not contribute to $\mtm^I$,
the primary constraints are
\beq
\psi_1^K=\mtm^K+2\pl_I\crd^{KI}\,. \eeq

The Hamiltonian is
\beq\label{Hamiltonian}\begin{split}
\Ham=&\frac12\left(\mtm_{IJ}\mtm^{IJ}-\frac1{\NN}\tr{\mtm}{I}\tr{\mtm}{J}\right)+
\frac2{\NN}\tr{\mtm}{I}\pl_J\crd^J+\beta_K\left(\mtm^K+2\pl_I\crd^{KI}\right)+\\
+&\frac12\left(\pl_I\crd_{JK}\pl^I\crd^{JK}+\pl_I\tr{\crd}{J}\pl^I\tr{\crd}{K}\right)
-\pl_I\crd_K\pl^I\crd^K+\\-
&\frac{d}{\NN}\pl^I\crd_I\pl^J\crd_J-\pl^I\crd_{IJ}\pl_K\crd^{KJ}\,,\end{split}
\eeq
where $\beta_K$ are Lagrange multipliers.

 Secondary and ternary constraints $\psi^K_2$, $\psi_3$
result from  Poisson brackets $[\ ,\ ]$ with the Hamiltonian
(\ref{Hamiltonian})\beq
[\psi^K_1,\Ham]=\psi^K_2=-\Delta\crd^K-\pl^K\pl_J\crd^J+\pl_J\mtm^{JK},\eeq
\beq\label{psi3}
[\psi^K_2,\Ham]=\pl^K\psi_3,\qquad
\psi_3=\Delta\tr{\crd}{I}-\pl_I\pl_J\crd^{IJ}\,,
\eeq
where $\Delta\equiv\pl^A\pl_A$.

The further commutation of the ternary constraints produces no new
constraints, so that $\psi_A$ (A=1,2,3) form the complete list.
All constraints are first class
\beq
[\psi_A,\psi_B]=0,\qquad A,B=1,2,3.
\eeq

In this analysis the kernel of the operator $\pl_K$ is assumed to be
trivial so that
 $\pl_K \psi_3=0 $ is equivalent to $\psi_3 =0$. Note however that
 the two conditions  may be different in a topologically nontrivial
 situation (say, for the torus compactification) differently accounting some
 discrete degrees of freedom. Note also that $\psi_3$ is just
 the linearized first-class constraint associated with $g^{00}$
 in usual Hamiltonian gravity \cite{dirham}.

 The number of physical degrees of freedom (PDoF) is  \beq
\mbox{PDoF}=\frac{(d+2)(d-1)}{2}-2(d-1)-1=\frac{d(d-3)}2=R(2,d-2)=2|_{d=4},\eeq
where $R(s,d)$ is the dimension of a rank-$s$ symmetric traceless
tensor
\beq
R(s,d)=\frac{(d+2s-2)(d+s-3)!}{(d-2)!s!}.
\eeq

Let us stress that the reason why the gauge fixed model under consideration turns
out to be equivalent to the Pauli-Fierz model is just that the
ternary constraint $\psi_3$ appears in (\ref{psi3}) under the operator
$\pl^K$, which, in turn, is the consequence of the Lorentz invariance of the chosen
gauge. In the temporary gauge electrodynamics example the equation analogous
to (\ref{psi3}) is (\ref{0div}) which is not a constraint however.

\subsection{Spin three}

Let us  consider a spin three massless field, as a simplest HS
example. We use the following notation for the space-like
projections of $\phi_{\mu\nu\lambda}$:
$\crd_{ABC}\equiv\phi_{ABC}$, $\crd_{AB}\equiv\phi_{AB0}$,
$\crd_{\phantom{B}BA}^{B}\equiv\phi_{A00}$,
$\crd_{\phantom{B}B}^{B}\equiv\phi_{000}$ and $\mtm_{ABC}$,
$\mtm_{AB}$ for the corresponding momenta.

The Hamiltonian that results from the action (\ref{TrAction}) has
the form
\beq\begin{split} \Ham&=\frac12{\mtm_{ABC}}^2-\frac3{2d}{\mtm_{\phantom{B}BA}^{B}}^2+\frac1{4(d-1)^2}{\mtm^{\phantom{B}B}_{B}}^2+
\frac6d\mtm^{\phantom{B}BA}_{B}\pl^C \crd_{AC}+\frac{3d}{2(d-1)^2}\mtm^{\phantom{B}B}_{B}\pl_A \crd^{\phantom{C}CA}_{C}+\\
&+ \frac12(\pl_A\crd_{BCD})^2-\frac32(\pl^A\crd_{ABC})^2+
\frac32(\pl_A\crd_{\phantom{B}BC}^{B})^2+\frac{3(d^2+4d-2)}{4(d-1)^2}(\pl^A\crd_{\phantom{B}BA}^{B})^2+\\&
-\frac32(\pl_A\crd_{BC})^2-\frac{3(d+2)}d(\pl^A\crd_{AB})^2-\frac12(\pl_A\crd^{\phantom{B}B}_{B})^2+\widetilde{\beta^{AB}}(\mtm_{AB}+3\pl^C\crd_{ABC}),
\end{split}
\eeq
where tilde\ \ $\widetilde{ }$\ \ denotes the traceless part; for
example $\widetilde{\crd_{AB}}\equiv\crd_{AB}-\frac1{(d-1)}
\delta_{AB}\crd^{C}_{\phantom{C}C}$. There are four generations of
constraints in this case:
\beq
{\widetilde{\psi_1^{AB}}}={\widetilde{\mtm^{AB}}}+3\widetilde{{\pl_C\crd^{ABC}}},
\eeq
\beq \widetilde{\psi^{AB}_2}=\frac{1}{3}[\widetilde{\psi^{AB}_1},\Ham]=\widetilde{\pl_C\mtm^{ABC}}-\widetilde{\Delta\crd^{AB}}-\widetilde{\pl^{(A}\pl_C\crd^{B)C}},\eeq
\beq\begin{split} &\qquad\qquad\qquad\qquad
\qquad[\widetilde{\psi^{AB}_2},\Ham]=
\widetilde{\pl^{(A}\psi^{B)}_3}, \\
&\psi^A_3=2\pl_B\pl_C\crd^{ABC}-2\Delta\crd^{\phantom{B}BA}_{B}-\frac1{(d-1)}\pl^A\mtm^{\phantom{B}B}_{B}-\frac{(d+2)}{(d-1)}\pl^A\pl_C\crd^{\phantom{B}BC}_{B},\end{split}\eeq
\beq\begin{split} &[\psi^{A}_3,\Ham]-\pl_B\psi^{AB}_2=\frac1{(d-1)}\pl^{A}\psi_4, \\&\psi_4=-d\Delta\crd^{\phantom{B}B}_{B}+\pl_A\mtm^{\phantom{B}BA}_{B}-2\pl_A\pl_B\crd^{AB}.\end{split}\eeq
Being first-class, the constraints
$\widetilde{\psi_1^{AB}}$, $\widetilde{\psi_2^{AB}}$,
$\psi_3^{A}$, $\psi_4$ along with the tracelessness conditions
 imply $R(3,d-2)$ physical degrees of freedom, which is two in $d=4$.

Let us compare these results with the Hamiltonian analysis of the
Fronsdal's formulation of spin three which goes as follows. The
primary constraints of Fronsdal's theory of massless spin three
$\widetilde{\psi_1^{AB}}$, $\psi_1^A$, $\psi_1$ are associated
with those components of the spin three field that carry index $0$
thus having a time derivative of the gauge parameter in their
transformation law. These generate the secondary first-class
constraints $\widetilde{\psi_2^{AB}}$, $\psi_2^A$, $\psi_2$ that
results in $R(3,d)+R(1,d)-2 R(2,d-1)-2
R(1,d-1)-2R(0,d-1)=R(3,d-2)=2|_{d=4}$ degrees of freedom.

In the traceless formulation considered here,
the constraints $\psi_1^A$ and
$\psi_1$ are absent, whereas $\psi_2^A$ and $\psi_2$ re-appear as
the constraints of third and fourth generation, $\psi_3^A$ and
$\psi_4$, respectively. As a result, compared to the Fronsdal
formulation, the deficit of first-class constraints equals
exactly to the deficit of field components so that the
 number of degrees of freedom remains unchanged. The counting
 of degrees of freedom for higher spins is analogous.

\subsection{Higher spins}\label{Higher Spins}
It is well-known (see \cite{Gitman:2004zf} for the formal proof)
that a number of first-class constraints equals to the number of
gauge parameters independent on a Cauchy surface assuming that
different time derivatives $\xi$, $\dot{\xi}$, $\ddot{\xi}$ are
independent on a Cauchy surface. Let us use this fact\footnote{We
acknowledge with gratitude that the idea of this analysis was
communicated to us by I.Tyutin} to count a number of first-class
constraints for a massless spin-$s$ field. Having decomposed
$\xi_{\mu_1...\mu_{s-1}}$ as
\beq\xi_{\underbrace{{0...0}}_{k}A(s-1-k-m)}=\sum_{i=0}^{i=s-k-1}\textstyle\frac{(s-k-1)!}{i!(s-k-i-1)!}
\pl_{(A_1}...\pl_{A_i}\xi^{k,i}_{A_{i+1}...A_{s-k-i-1})},\eeq with
$\pl^B\xi^{k,i}_{BA_{2}...A_{s-k-1}}=0$, the tracelessness
condition $\xi^{\nu}_{\phantom{\nu}\nu\mu_3...\mu_{s-1}}=0$ and
(\ref{divLessCovConstr}) acquire the form \beq\begin{split}
&\pl_0\xi^{k+1,i}_{A(s-k-i-2)}=\Delta\xi^{k,i+1}_{A(s-k-i-2)},\\&\xi^{k+2,i}_{A(s-k-i-3)}=\Delta\xi^{k,i+2}_{A(s-k-i-3)}+\xi^{k,i}_{A(s-k-i-3)BB}\delta^{BB}.\end{split}\eeq
The first equation allows us to express $\xi^{k,m}$ with $m>0$ via
time-derivatives of $\xi^{k+m,0}$ as \beq
\xi^{k,m}_{A(s-k-m-2)}=\frac{(\pl_0)^m}{\Delta^m}\xi^{k+m,0}_{A(s-k-m-2)}\,,
\eeq whereas the second one states that the trace of $\xi^{k,0}$
 is expressed via $\xi^{k+2,0}$ as
\beq
\left(1-\frac{(\pl_0)^2}{\Delta}\right)
\xi^{k+2,0}_{A(s-k-3)}=\xi^{k,0}_{A(s-k-3)BB}\delta^{BB}\,.
\eeq
As a result, the traceless components of $\xi^{k,0}_{A_1...A_{s-k-1}}$ with
$k=0...(s-1)$ remain the only independent parameters. The number of
independent gauge parameters that appear in (\ref{MasslessGauge})
with $(\pl^0)^{r}$  is $R(s-r,d-1)$ for $r\geq1$ and $R(s-1,d-1)$ for
$r=0$. This gives the correct number of physical degrees of
freedom
\beq
\mbox{PDoF}=R(s,d)-2R(s-1,d-1)-\sum_{k=0}^{k=s-2}R(k,d-1)\equiv
R(s,d-2),\eeq which is the dimension of the spin-$s$ irreducible
representation of the massless little group \mls, which is two in
$d=4$.

Since first-class constraints associated with the gauge
transformations generated by the parameters carrying $r$ time
derivatives appear as constraints of $r^{th}$ generation, this
analysis also explains why in the transverse formulation of a
spin-$s$ field, first-class constraints appear up to the
$(s+1){}^{th}$ generation.

Note also that the two models are equivalent in the topologically
trivial situation with invertible space-like derivatives simply
because the partial gauge fixing that reduces the Fronsdal model
to the transverse gauge invariant can be interpreted as being of
Stueckelberg type with respect to the components of the gauge
parameters contracted with the space-like derivatives in the
transversality  condition (\ref{divLessCovConstr}). The reason why
this is not true
 in the example of electrodynamics in the temporary gauge
is that  the condition (\ref{de}) contains only time derivative.

To conclude,
the formulation of massless fields in terms of
traceless tensors is equivalent to the original Fronsdal
formulation in the topologically trivial situation although
it may be different otherwise.

\section*{Acknowledgements}  We grateful to A.Barvinsky and especially
to I. Tyutin for useful discussions of the related aspects of
Hamiltonian dynamics, and to M.~Tsulaia for bringing to our
attention \cite{Pashnev:1998ti}.
 The work was supported in part by
grants RFBR No. 05-02-17654, LSS No. 1578.2003-2, 4401.2006.2 and
INTAS No.  05-7928. The work of E.S. was also supported by the
Landau Scholarship and by the Scholarship of Dynasty foundation.

\end{document}